\documentclass[twocolumn]{aastex631}
\usepackage{amsmath}

\begin{document}

\newcommand{\vdag}{(v)^\dagger}
\newcommand\aastex{AAS\TeX}
\newcommand\latex{La\TeX}

\newcommand\JC{70}
\newcommand\JM{87}

\newcommand\HM{83}
\newcommand\HC{66}
\newcommand\JHM{35}
\newcommand\JHC{32}

\newcommand\MC{94}
\newcommand\MM{114}

\newcommand\trgbswa{20.62}
\newcommand\trgbswb{19.27}
\newcommand\trgblwa{18.98}
\newcommand\trgblwb{18.80}
\newcommand\dtrgbswa{0.03}
\newcommand\dtrgbswb{0.05}
\newcommand\dtrgblwa{0.07}
\newcommand\dtrgblwb{0.05}

\newcommand\trgbma{19.79}
\newcommand\trgbmb{19.57}
\newcommand\trgbmc{19.26}
\newcommand\dtrgbma{0.04}
\newcommand\dtrgbmb{0.04}
\newcommand\dtrgbmc{0.04}

\newcommand{\todo}[1]{\textcolor{red}{TODO: #1}}

\title{The JWST Resolved Stellar Populations Early Release Science Program VI. Identifying Evolved Stars in Nearby Galaxies}

\correspondingauthor{Martha L.\ Boyer}
\email{mboyer@stsci.edu}

\author[0000-0003-4850-9589]{Martha L.\ Boyer}
\affiliation{Space Telescope Science Institute, 3700 San Martin Drive, Baltimore, MD 21218, USA}

\author[0000-0002-9300-7409]{Giada Pastorelli}
\affiliation{Padova Astronomical Observatory, Vicolo dell'Osservatorio 5, Padova, Italy}

\author[0000-0002-6301-3269]{L\'eo Girardi}
\affiliation{Padova Astronomical Observatory, Vicolo dell'Osservatorio 5, Padova, Italy}

\author[0000-0002-9137-0773]{Paola Marigo}
\affiliation{Department of Physics and Astronomy G. Galilei, University of Padova, Vicolo dell’Osservatorio 3, I-35122, Padova, Italy}

\author[0000-0001-8416-4093]{Andrew E. Dolphin}
\affiliation{RTX, 1151 E. Hermans Road, Tucson, AZ 85756, USA}
\affiliation{Steward Observatory, University of Arizona, 933 N. Cherry Avenue, Tucson, AZ 85719, USA}

\author[0000-0001-5538-2614]{Kristen.~B.~W. McQuinn}
\affiliation{Rutgers University, Department of Physics and Astronomy, 136 Frelinghuysen Road, Piscataway, NJ 08854, USA}

\author[0000-0002-8092-2077]{Max~J.~B. Newman}
\affiliation{Rutgers University, Department of Physics and Astronomy, 136 Frelinghuysen Road, Piscataway, NJ 08854, USA}

\author[0000-0002-1445-4877]{Alessandro Savino}
\affiliation{Department of Astronomy, University of California, Berkeley, CA 94720, USA}

\author[0000-0002-6442-6030]{Daniel R. Weisz}
\affiliation{Department of Astronomy, University of California, Berkeley, CA 94720, USA}

\author[0000-0002-7502-0597]{Benjamin F. Williams}
\affiliation{Department of Astronomy, University of Washington, Box 351580, U.W., Seattle, WA 98195-1580, USA}

\author[0000-0003-2861-3995]{Jay Anderson}
\affiliation{Space Telescope Science Institute, 3700 San Martin Drive, Baltimore, MD 21218, USA}

\author[0000-0002-2970-7435]{Roger E. Cohen}
\affiliation{Rutgers University, Department of Physics and Astronomy, 136 Frelinghuysen Road, Piscataway, NJ 08854, USA}

\author[0000-0001-6464-3257]{Matteo Correnti}
\affiliation{INAF Osservatorio Astronomico di Roma, Via Frascati 33, 00078, Monteporzio Catone, Rome, Italy}
\affiliation{ASI-Space Science Data Center, Via del Politecnico, I-00133, Rome, Italy}

\author[0000-0003-0303-3855]{Andrew A. Cole}
\affiliation{School of Natural Sciences, University of Tasmania, Private Bag 37, Hobart, Tasmania 7001, Australia}

\author[0000-0002-7007-9725]{Marla C. Geha}
\affiliation{Department of Astronomy, Yale University, New Haven, CT 06520, USA}

\author[0000-0002-5581-2896]{Mario Gennaro}
\affiliation{Space Telescope Science Institute, 3700 San Martin Drive, Baltimore, MD 21218, USA}
\affiliation{The William H. Miller {\sc III} Department of Physics \& Astronomy, Bloomberg Center for Physics and Astronomy, Johns Hopkins University, 3400 N. Charles Street, Baltimore, MD 21218, USA}

\author[0000-0002-3204-1742]{Nitya Kallivayalil}
\affiliation{Department of Astronomy, The University of Virginia, 530 McCormick Road, Charlottesville, VA 22904, USA}

\author[0000-0001-6196-5162]{Evan N.\ Kirby}
\affiliation{Department of Physics and Astronomy, University of Notre Dame, 225 Nieuwland Science Hall, Notre Dame, IN 46556, USA}

\author[0000-0002-4378-8534]{Karin M. Sandstrom}
\affiliation{Department of Astronomy \& Astrophysics, University of California San Diego, 9500 Gilman Drive, La Jolla, CA 92093, USA}

\author[0000-0003-0605-8732]{Evan D. Skillman}
\affiliation{University of Minnesota, Minnesota Institute for Astrophysics, School of Physics and Astronomy, 116 Church Street, S.E., Minneapolis,
MN 55455, USA}

\author[0000-0001-9061-1697]{Christopher T. Garling} 
\affiliation{Department of Astronomy, The University of Virginia, 530 McCormick Road, Charlottesville, VA 22904, USA}

\author[0000-0002-3188-2718]{Hannah Richstein}
\affiliation{Department of Astronomy, The University of Virginia, 530 McCormick Road, Charlottesville, VA 22904, USA}

\author[0000-0003-1634-4644]{Jack T. Warfield}
\affiliation{Department of Astronomy, The University of Virginia, 530 McCormick Road, Charlottesville, VA 22904, USA}







\begin{abstract}

We present an investigation of evolved stars in the nearby star-forming galaxy WLM, using NIRCam imaging from the JWST resolved stellar populations early-release science (ERS) program. We find that various combinations of the F090W, F150W, F250M, and F430M filters can effectively isolate red supergiants (RSGs) and thermally-pulsing asymptotic giant branch (TP-AGB) stars from one another, while also providing a reasonable separation of the primary TP-AGB subtypes: carbon-rich C-type stars and oxygen-rich M-type stars. The classification scheme we present here agrees very well with the well-established Hubble Space Telescope (HST) medium-band filter technique. The ratio of C to M-type stars (C/M) is 0.8$\pm$0.1 for both the new JWST and the HST classifications, which is within one sigma of empirical predictions from optical narrow-band CN and TiO filters. The evolved star colors show good agreement with the predictions from the PARSEC$+$COLIBRI stellar evolutionary models, and the models indicate a strong metallicity dependence that makes stellar identification even more effective at higher metallicity. However, the models also indicate that evolved star identification with NIRCam may be more difficult at lower metallicies. We test every combination of NIRCam filters using the models and present additional filters that are also useful for evolved star studies. We also find that $\approx$90\% of the dusty evolved stars are carbon-rich, suggesting that carbonaceous dust dominates the present-day dust production in WLM, similar to the findings in the Magellanic Clouds. These results demonstrate the usefulness of NIRCam in identifying and classifying dust-producing stars without the need for mid-infrared data.

\end{abstract}


\section{Introduction}
The thermally-pulsing Asymptotic Giant Branch (TP-AGB) stage is the end-phase of stellar evolution for $\sim$1--8~$M_{\odot}$ stars, marked by complex physical processes such as pulsation, mass loss, hot bottom burning, and dredge up \citep{Habing2004,Karakas2017}. TP-AGB stars are relatively rare due to the short lifetimes ($\sim0.5$--$20$~Myr) of this phase, yet they have an outsized impact on the evolution of their host galaxies owing to their production of metals and dust. Their luminosities are also extremely high, affecting interpretations of galaxy observations. 

Stellar models have historically struggled to reproduce TP-AGB observables owing to both the complexity of this stage of evolution and to a scarcity of high-quality calibration data. Recent progress has been made in calibrating the TP-AGB phase \citep[e.g.,][]{Rosenfield+2014, Rosenfield+2016} using data of nearby galaxies, but many of these efforts were unable to separate different TP-AGB subtypes in the data, which limits the scope of the calibration. The only galaxies with high quality, complete TP-AGB calibration data (including stellar subtypes) are the Magellanic Clouds, for which there is an abundance of near- to mid-infrared (IR) data. As a result, most models are anchored to the Magellanic Clouds \citep[e.g.,][]{Pastorelli+2019,Pastorelli+2020}, leaving the metallicity dependencies poorly constrained.  
These model uncertainties propagate when TP-AGB stars are present in observations that are used to derive various galaxy properties.

The presence of TP-AGB stars can make precise measurements of the tip of the red giant branch (TRGB) difficult since TP-AGB stars can overlap with RGB stars.  In addition, fitting a stellar population to derive model-dependent parameters such as age, star-formation history (SFH), or metallicity can result in highly inaccurate estimates if the TP-AGB stars are not excluded or fully accounted for \citep{conroy+2009, Baldwin+2018}. Moreover, failure to separate oxygen-rich M-type AGB stars from carbon-rich C-type AGB stars can also potentially affect distance estimates measured from period-luminosity (P-L) diagrams and the `JAGB' method \citep{Freedman+2020, Lee+2023a}.

Classifying TP-AGB stars can also provide useful galaxy diagnostics. For example, the ratio of C- to M-type stars (C/M) can provide information about a galaxy's metallicity and/or age \citep[e.g.,][]{Groenewegen2006b, Boyer+2019}.  Also, since each subtype produces a different type of dust (carbonaceous or siliceous), separating them allows for a better understanding of the source of a galaxy's interstellar dust: AGB stars, supernovae, or grain growth in the interstellar medium itself.

Yet, identifying TP-AGB stars, and especially classifying them into their main subtypes (M- and C-type) and separating them from other evolved stars like red supergiants (RSGs), is challenging without spectroscopy and pulsation information. While a fraction of TP-AGB stars is always found below the TRGB because of the luminosity dips of their thermal pulse cycles, in the optical wavelengths $<$1~\micron\ this fraction increases due to molecular absorption in their atmospheres and/or to variability and extinction from circumstellar dust \citep{Boyer+2019}. This is less of an issue in the near-IR, where molecular absorption is not as strong, pulsation amplitudes are smaller, and extinction is weaker. However, clean and reliable separation of M- and C-type AGB stars in the near-IR has so far proven impossible with broadband filters \citep[e.g.,][]{Girardi+2010, Boyer+2013, Boyer+2015, Rosenfield+2014}. 

Recent studies have shown that medium-band filters on HST's Wide Field Camera 3 (WFC3/IR) cleanly sample key molecular features and can thus effectively and efficiently identify TP-AGB stars {\em and} classify them into their subtypes. However, given the limitations in HST's resolution and sensitivity (and its finite lifetime), finding effective filter combinations with JWST's Near-Infrared Camera \citep[NIRCam;][]{Rieke+2005,Gardner+2023,Rigby+2023,Rieke+2023} is prudent. 

To investigate JWST's ability to photometrically classify evolved stars, we target the nearby dwarf irregular galaxy Wolf--Lundmark--Melotte (WLM)\@ as part of the JWST Stellar Populations Early Release Science (ERS) program \citep{Boyer+2022, Warfield+2023, Weisz+2023, McQuinn+2024}.  \citet{wolf1909} discovered the galaxy on photographic plates, and \citet[][along with K.~Lundmark]{melotte1926} re-discovered the galaxy and noted its similar appearance to NGC~6822, the nearest dwarf irregular galaxy.  WLM is seen nearly edge-on to its rotating disk of gas and stars \citep{leaman12}.  Like most dwarf irregular galaxies, WLM has ongoing star formation \citep[$0.008~M_{\sun}~{\rm yr}^{-1}$][]{mondal18}.  It also has a star formation history weighted toward recent ages, with the majority of star formation occurring in the last 5~Gyr \citep{weisz14,Albers+2019,McQuinn+2024}.  WLM's stellar metallicity distribution is typical for dwarf galaxies of its mass, with a mean iron abundance of ${\rm [Fe/H]} = -1.27$ and a shallow radial metallicity gradient \citep{Leaman+2009,leaman13}.

One goal of our ERS program is to demonstrate whether a filter combination that is popular for many galaxy studies can enable accurate AGB and RSG identification and classification. In this paper, we explore NIRCam data for WLM and compare the evolved star population to Hubble observations and to that predicted by stellar models. In \S\ref{sec:data} we describe the observations and data reduction. In \S\ref{sec:analysis} we outline our selection of TP-AGB stars and other objects. In \S\ref{sec:disc} we discuss our final selection recommendation and galaxy properties that might affect it, such as the metallicity. 


\section{Data}
\label{sec:data}

The NIRCam images of WLM\footnote{\dataset[DOI: 10.17909/71kb-ga31]{https://doi.org/10.17909/71kb-ga31}} were observed as part of ERS program 1334 (PI Weisz) in July 2022 using the F090W and F150W filters in the short wavelength (SW) channel and the F250M and F430M filters in the long wavelength (LW) channel. See \citet{Weisz+2023} for full details about the observations.  The exposure setup included 8 groups with the \texttt{MEDIUM8} readout pattern with 7--9 integrations and 4 subpixel dithers (\texttt{4-POINT-MEDIUM-WITH-NIRISS}). The total exposure time was $\approx$24--30~ks. NIRCam's Module A covers a central field in WLM, while Module B covers a field to the south. 

The data were processed with JWST Pipeline version 1.11.4, with CRDS context  jwst\_p1147.pmap, using the default parameters in the pipeline through Stage-2 processing. Photometry was performed using the new NIRCam module for \texttt{DOLPHOT} \citep{Dolphin2000, Dolphin2016, Weisz+2023}, using parameters described by \citet{Weisz+2024}.  The 50\% photometric completeness limit is $\gtrsim$28~mag in the broadband filters and $\sim$25~mag in the medium band filters, both of which are several magnitudes fainter than the TRGB\@. The photometric uncertainties are within 0.01~mag. 
The photometry was culled following \citet{Warfield+2023}, see \S\ref{sec:contam}. A handful of stars in our images are saturated, so it is possible that a few of the most extreme examples of TP-AGB stars are not included in our catalog, but we expect the TP-AGB population is very near fully complete in the NIRCam FOV\@.

\begin{figure}
\includegraphics[width=0.52\columnwidth, angle=90]{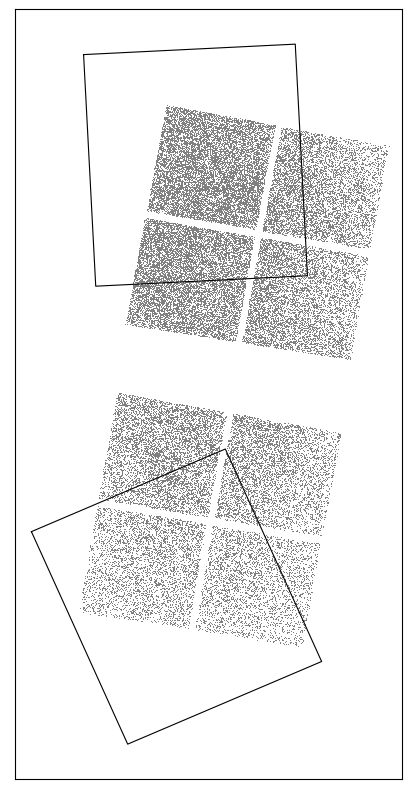}
\caption{Map of the SW NIRCam point sources (black dots), with the medium-band HST WFC3/IR footprints from GO-16162 overlaid in black. North is to the left and east is down. Each NIRCam module (4 SW detectors with $\approx$4\arcsec small gaps in between) is 2\farcm2$\times$2\farcm2.
\label{fig:map}}
\end{figure}

Our analysis also includes Hubble Space Telescope (HST) data from the Local UltraViolet and Infrared Treasury (LUVIT) program (K. Gilbert et al., in prep, and M. Boyer et al., in prep). That program includes imaging with the Advanced Camera for Surveys (ACS), Wide Field Camera 3 (WFC3), and Wide-Field and Planetary Camera 2 (WFPC2) instruments in at least nine filters from F275W to F160W, obtained as part of programs GO-15275 (PI: K. Gilbert) and GO-16162 (PI: M. Boyer).  Photometry on those data was performed using \texttt{DOLPHOT} and the photometry pipeline initially developed for the Panchromatic Hubble Andromeda Treasury (PHAT) program \citep{Dalcanton+2012, Williams+2014}. Here, we focus mainly on the IR medium-band filters from GO-16162: F127M, F139M, and F153M, which are particularly useful for separating M-type from C-type AGB stars (\S\ref{sec:HSTagb}). Figure~\ref{fig:map} shows the overlap between the JWST and HST fields of view.

Throughout our analysis, we assume a foreground extinction of $A_V = 0.10$ based on Galactic dust maps \citep{Schlegel+1998, Schlafly+2011}, and adopt the $A_\lambda$ values using the \citet{Cardelli+1989} $+$ \citet{ODonnell1994} extinction curve with $R_V = 3.1$ and the post-launch JWST filters transmissions updated in Nov 2022.\footnote{Derived using the \href{http://stev.oapd.inaf.it/cmd}{CMD v3.2 website}}  

\section{Analysis}
\label{sec:analysis}

\subsection{TRGB Estimates}
\label{sec:trgb}

At most wavelengths, TP-AGB stars are usually brighter than the TRGB, making the TRGB a key constraint for identifying AGB stars.  However, multiple factors can push TP-AGB stars fainter than this limit.  Thermal pulses, variability and/or strong molecular absorption in the atmosphere can cause these stars to periodically dip below the TRGB at some wavelengths, especially in optical and near-IR bands where these effects cause the largest brightness changes. Circumstellar dust can also cause a highly evolved TP-AGB star to appear fainter than the TRGB in bluer bands. For all of these reasons, including the TRGB at longer wavelengths in the selection criteria results in a more complete TP-AGB sample.

To measure the TRGB in each band, we follow the procedure in \citet{Mendez+2002}. We create a Gaussian-smoothed luminosity function for each filter and apply an edge-detection Sobel kernel of the form $[-2, 0, +2]$. We repeat this with 1000 Monte Carlo bootstrap resampling trials, each time adding random 1-$\sigma$ noise to the photometry and randomly varying the starting magnitude of the luminosity function bins. The final TRGB magnitude and its uncertainty are the peak and standard deviation of a Gaussian function fit to the distribution of TRGBs measured in the trials; these are listed in Table~\ref{tab:trgb}. With our adopted distance modulus of 24.93~mag \citep{Albers+2019}, the absolute magnitude of the F090W TRGB is $M_{\rm TRGB}^{\rm F090W} = -4.31$~mag, within 0.05~mag of the F090W estimate recently reported by \citet{Anand+2024}. 


\begin{deluxetable}{lcc}
\tablecaption{Measured TRGBs \label{tab:trgb}}
\tablehead{
\colhead{Filter} & \colhead{TRGB}& \colhead{$\sigma$TRGB} \\
\colhead{} & \colhead{(mag)} &
\colhead{(mag)}}
\startdata 
\multicolumn{3}{c}{---JWST NIRCam---}\\
F090W & \trgbswa & \dtrgbswa\\
F150W & \trgbswb & \dtrgbswb\\
F250M & \trgblwa & \dtrgblwa\\
F430M & \trgblwb & \dtrgblwb\\
\tableline
\multicolumn{3}{c}{---HST WFC3/IR---}\\
F127M & \trgbma & \dtrgbma\\
F139M & \trgbmb & \dtrgbmb\\
F153M & \trgbmc & \dtrgbmc\\
\enddata
\end{deluxetable}

\subsection{Selecting TP-AGB stars with HST filters}
\label{sec:HSTagb}

The medium-band HST filters overlap with the H$_2$O and CN+C$_2$ features in M-type and C-type TP-AGB spectra, respectively. A combination of the F127M, F139M, and F153M filters are therefore well-suited to isolating these types of TP-AGB stars \citep{Boyer+2013, Boyer+2017, Boyer+2019, Goldman+2019, Jones+2023}. The LUVIT HST fields overlap with the ERS JWST fields (Fig.~\ref{fig:map}), so we can use the HST classifications to guide TP-AGB classification in the ERS filters. 

\begin{figure}
    \centering
    \includegraphics[width=\columnwidth]{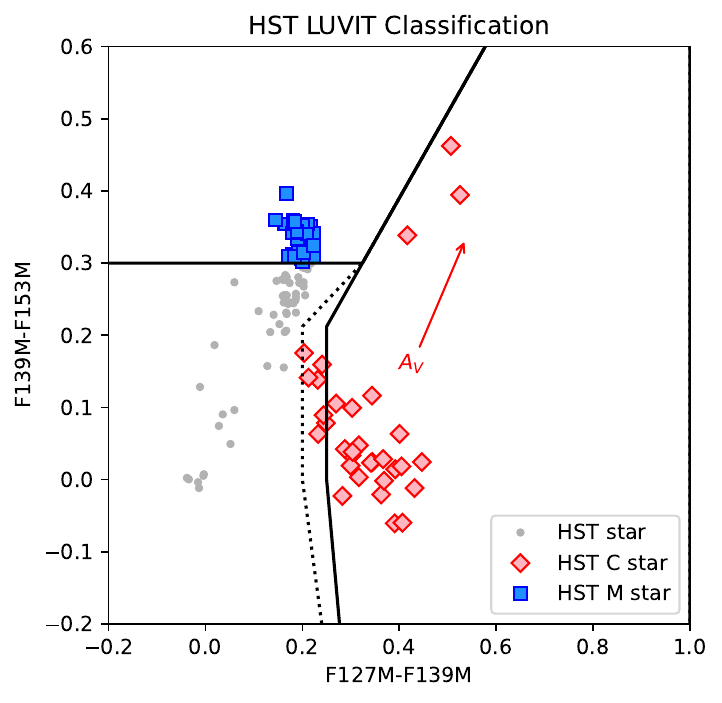}
    \caption{HST medium-band color-color diagram showing the separation between C-type and M-type TP-AGB stars in the overlapping HST$+$JWST fields (Fig.~\ref{fig:map}). All points in this diagram are brighter than the TRGB in at least one of the HST filters. Grey points are dominated by foreground, main sequence, and RSGs \citep{Boyer+2013, Boyer+2017}. The solid line shows the boundaries derived by \citet{Boyer+2019}, which we adjust slightly blueward (dotted line) to include a handful of additional stars that appear to be an extension of the branch of C-type stars. The red arrow shows the direction and magnitude of circumstellar extinction for $E(J-K_{\rm s}) = 1$ mag for a mix of 70\% amorphous carbon and 30\% SiC dust \citep{Groenewegen2006a}.
    \label{fig:hstccd1}}
\end{figure}

\begin{figure*}
\includegraphics[width=\textwidth]{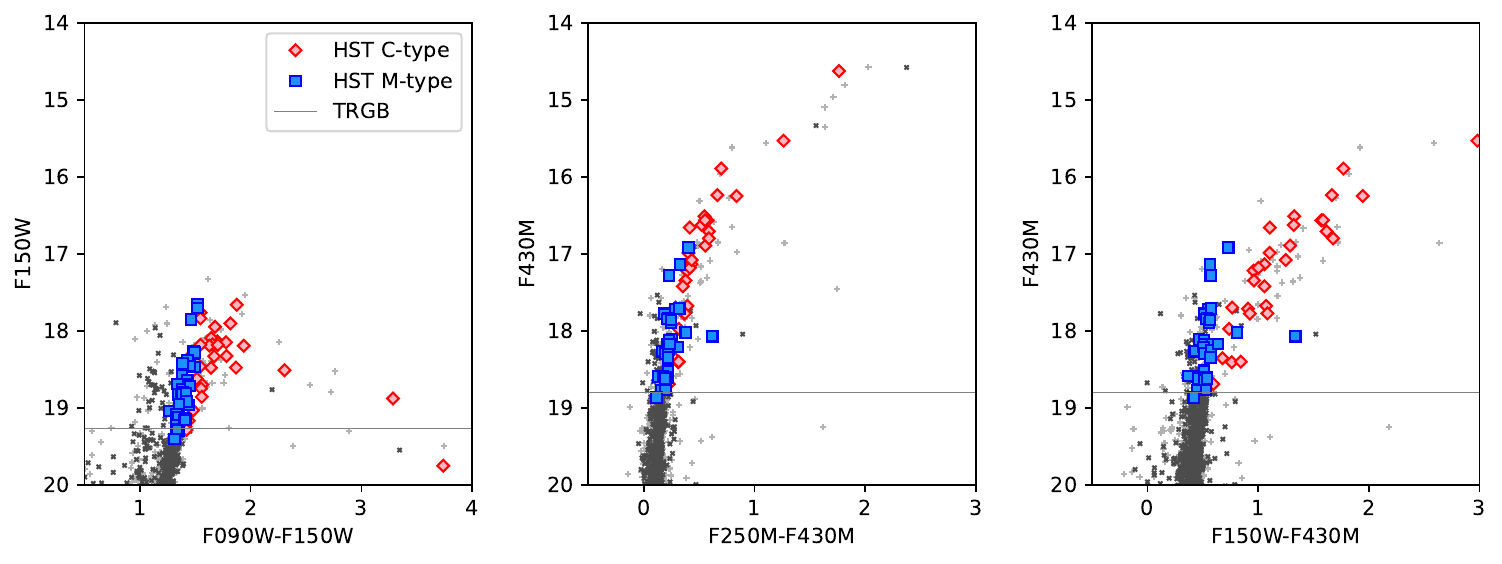}
\caption{JWST CMDs with HST-selected TP-AGB stars overplotted. Light grey points are all stars, and dark points are stars in the region where the WFC3/IR and NIRCam footprints overlap. Blue squares and red diamonds are stars classified using the HST medium-band photometry, which covers a smaller FOV inside the NIRCam footprint (Fig.~\ref{fig:map}). While the separation of the TP-AGB subtypes is poor when only considering the ERS SW (left) or LW (middle) bands in isolation, we find good separation by combining both (right). 
\label{fig:jwstcmds}}
\end{figure*}

In Figure~\ref{fig:hstccd1}, we show the \citet[][B19]{Boyer+2019} selection boundaries, which recover more M-type stars than the \citet[][B17]{Boyer+2017} boundaries. The original B17 boundaries are based on stellar atmosphere models from \citet{Aringer+2009, Aringer+2016}, which B19 adjusted slightly to better fit HST data of the M31 galaxy. The exact boundaries between stellar types may be slightly dependent on stellar metallicity and the age of the population. Here, we see that a handful of stars around $\rm{F127M}-\rm{F139M} = 0.25$~mag are excluded from the B19 C-type selection, despite appearing to be an extension of the C star branch. We therefore adjust the boundary slightly blueward to include these stars (dotted line in Fig.~\ref{fig:hstccd1}). Inspection of Aringer's library of C star spectra indicates that the F127M$-$F139M interval between 0.2 and 0.25~mag can be filled with C stars of relatively high effective temperature ($\sim$4000~K), which probably form only at low metallicities.


Figure~\ref{fig:jwstcmds} shows three NIRCam color-magnitude diagrams (CMDs) with the HST-selected TP-AGB stars plotted in red and blue. It is clear that, with our filters, a combination of SW and LW filters is needed for a robust separation of C- and M-type stars since the SW-only (left) and LW-only (middle) CMDs show the highest degree of overlap between the two types.


\subsection{Selecting TP-AGB stars with JWST filters}

\begin{figure}
\includegraphics[width=\columnwidth]{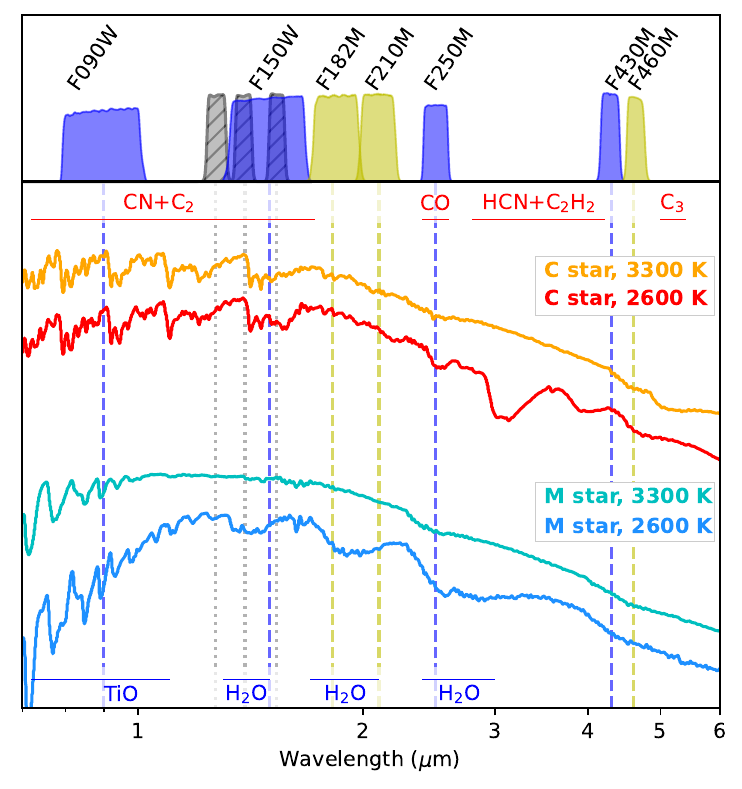}
\caption{Model stellar atmospheres (bottom panel) of C- and M-type stars from \citet{Aringer+2009, Aringer+2016}. The ERS NIRCam filters are shown in blue in the top panel. Medium-band HST WFC3/IR filters are shown in gray and hatched (F127M, F139M, F153M). Additional NIRCam filters discussed in \S\ref{sec:others} are plotted in yellow.  The dashed and dotted lines in the bottom panel indicate the center wavelengths of the filters. The most prominent molecular features are marked in the bottom panel for C-type stars (red) and M-type stars (blue).
\label{fig:filters}}
\end{figure}

The ERS SW filters (F090W and F150W) were selected to provide a good recovery of the WLM SFH \citep{McQuinn+2024}. The LW filters were observed simultaneously at no additional cost; we selected F250M and F430M for the LW channel, in part because the COLIBRI TP-AGB models \citep{Marigo+2013, Marigo+2017} predicted that they would provide a reasonable separation between TP-AGB stellar types when used together with the SW filters.  The two LW filters sample CO in the atmospheres of C-type stars and H$_2$O in M-type stars (Fig.~\ref{fig:filters}). The F430M filter is also sensitive to emission from circumstellar dust.  On the SW side, the F150W filter samples H$_2$O in M-type stars, while both filters sample CN$+$C$_2$ in C-type stars.  See Figure~\ref{fig:filters} for example model atmosphere spectra of C- and M-type TP-AGB stars from \citet{Aringer+2009, Aringer+2016}, which are used to calculate the bolometric corrections applied to TP-AGB stars in the COLIBRI tracks.  In this section, we use the HST classifications and the COLIBRI models to guide the JWST TP-AGB classification.

\begin{figure*}
    \centering
    \includegraphics[width=\textwidth]{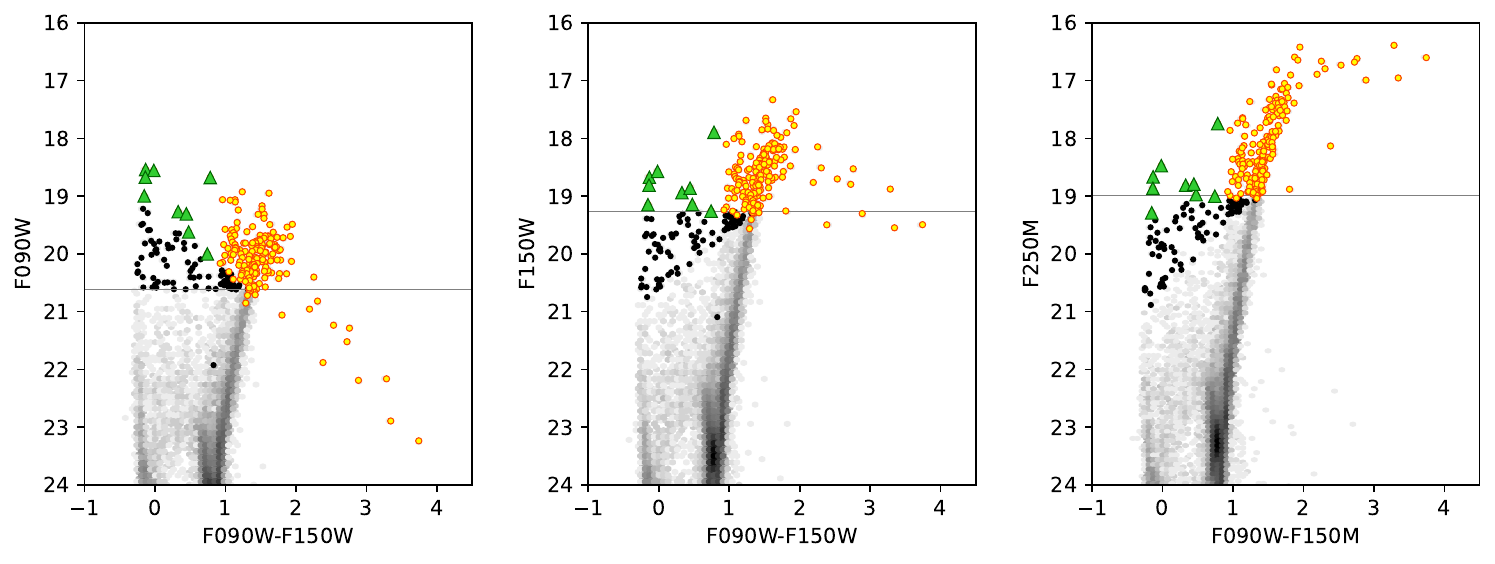}
    \caption{CMDs showing which stars are brighter/fainter than the TRGB at F090W, F150W, and F250M. Green and Yellow points are those that are brighter than the TRGB in only F150W or F250M. Green points are those with F090W$-$F150W $< 0.8$~mag and include contaminating main sequence and blue helium-burning stars. Yellow points are dominated by AGB stars and RSGs. Black points are remaining stars that are brighter than the TRGB in any of the four ERS filters, and include mostly contaminating sources (main sequence and blue helium-burning stars). Horizontal lines mark the TRGB in the F090W (left), F150W (middle), and F250M (right) filters. To maximize the evolved star sample while minimizing contaminating sources, our classification scheme starts with stars that are brighter than either F150W or F250M.
    \label{fig:trgbjust}}
\end{figure*}

\subsubsection{The effect of the JWST TRGB}
\label{sec:jwsttrgb}

The TRGB limit provides a magnitude cut in each band to eliminate RGB stars and other faint objects. Here, we investigate which TRGB limits provide the most complete sample of TP-AGB stars with minimal contamination. As mentioned in \S\ref{sec:trgb}, TP-AGB stars can sometimes fall below the TRGB due to variations during the pulsation cycle, molecular absorption, and/or circumstellar dust attenuation, with the strength of the effect depending on the wavelength. Figure~\ref{fig:trgbjust} shows that some dusty AGB stars fall below the F090W and F150W TRGBs (the yellow points with the reddest colors in the CMDs). In addition, a large fraction of blue contaminating stars, such as main sequence and blue He-burning stars, are brighter than the F090W TRGB. 

Both of these issues are mitigated by instead using a TRGB measured in longer wavelengths. The right panel in Figure~\ref{fig:trgbjust} shows that the F250M TRGB minimizes contamination from blue sources while also including dusty AGB stars. We therefore compile our starting TP-AGB sample by including stars that are either brighter than the F150W {\em or} F250M TRGBs. We exclude the F430M TRGB because very red contaminating sources such as unresolved background galaxies and embedded young stellar objects can be brighter than the TRGB at wavelengths redder than about 3~\micron\ \citep[e.g.,][]{Jones+2017}.

\begin{figure*}
\centering
\includegraphics[width=0.85\textwidth]{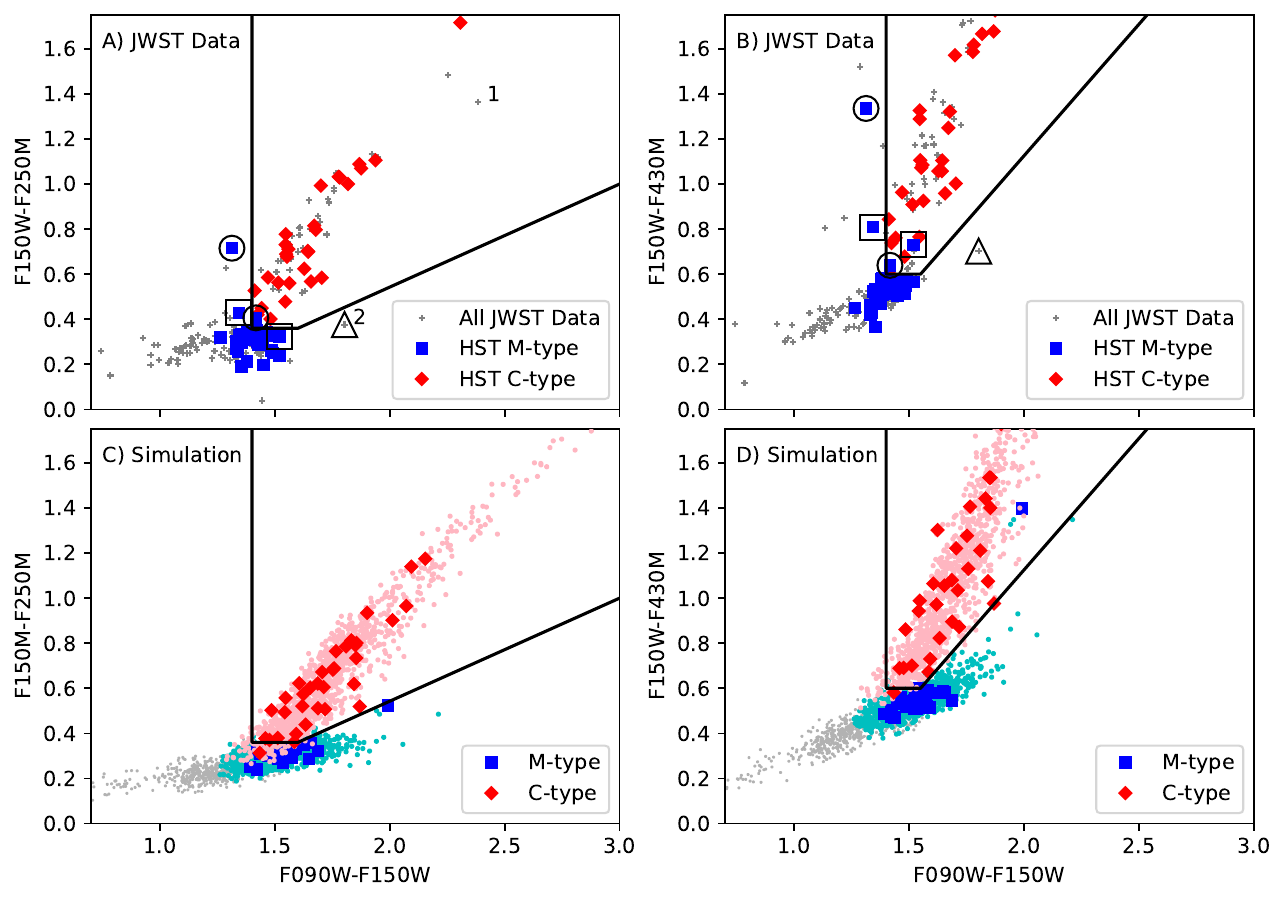}
\caption{Color-color diagrams used to classify C- and M-type TP-AGB stars. The upper panels (A and B) show the JWST data, while the lower panels (C and D) show simulations based on the COLIBRI models and the SFH from \citet{McQuinn+2024}, scaled to match the number of AGB stars in the HST$+$NIRCam footprint. In all panels, only stars that are brighter than either the F150W or F250M TRGBs are included. Red diamonds are C-type stars and blue squares are M-type stars, classified via HST colors (panels A and B) or identified via their model parameters (panels C and D). Light pink and cyan dots in the lower panels show the same simulation for 10$\times$ WLM's stellar mass, to illustrate more sparsely populated regions of these diagrams. The black lines mark the JWST color selections (Eqs.~\ref{eq:250} and \ref{eq:430}). Stars of interest are outlined in black circles, squares, and triangles, see text. In panel A, possible contaminants 1 and 2 are marked (see \S\ref{sec:contam}).
\label{fig:jwstccd}}
\end{figure*}


\subsubsection{JWST TP-AGB Classification}
\label{sec:jwstclass}

Based on the HST TP-AGB classifications for WLM, we find that the widest color separation between stellar types is in the F090W$-$F150W vs. F150W$-$F250M color-color diagram (CCD), shown in Figure~\ref{fig:jwstccd}, panel A. The CCD with F150W$-$F430M on the y-axis (panel B) is also promising, but there is more overlap between M-type stars (marked by large black squares or circles) and the C-type stars. This overlap is concerning for systems that may have a larger population of late-type M stars with deep H$_2$O absorption. In Panel A, these stars are more separated from the C star population. The black lines in these panels mark our best separation between AGB subtypes based on the HST classification, chosen to minimize cross-contamination between C- and M-type stars in these JWST colors:

\begin{multline}
{\rm F250M:}\\
(F090W-F150W, F150W-F250M)_{\rm HST{\text -}driven} \\
= [(1.4, 2.0), (1.4, 0.36), (1.6, 0.36), (3.0,1.0)], 
\label{eq:250}
\end{multline}

\begin{multline}
{\rm F430M:}\\
(F090W-F150W, F150W-F430M)_{\rm HST{\text -}driven} = \\
[(1.4,2.0), (1.4,0.6), (1.55,0.60), (2.75,2.0)].
\label{eq:430}
\end{multline}

These AGB star selections are supported by comparing to model predictions. Using the star formation history (SFH) from \citet{McQuinn+2024}, we simulate the population of WLM using the TRILEGAL code \citep{Girardi+2005} and the COLIBRI TP-AGB tracks from \citet{Pastorelli+2019, Pastorelli+2020}.
The bolometric corrections for AGB stars are based on the stellar spectral libraries by \citet{Aringer+2009} for C-rich stars, and \citet{Aringer+2016} for O-rich stars. 
The effect of circumstellar dust is taken into account following the approach of \citet{Marigo+2008} and adopting the following dust mixtures: amorphous carbon (85\%) and SiC  (15\%)  for  C-rich  stars,  and  silicates  for  O-rich stars \citep{Groenewegen2006a}. For non-TP-AGB stars, the simulation uses tracks from the PARSEC v1.2S database \citep{bressan12,tang14,chen15}. 

The COLIBRI tracks are calibrated with TP-AGB stars from the Small Magellanic Cloud (SMC) for $0.0005 \leq Z_{\rm i} \leq 0.006$ and from the Large Magellanic Cloud (LMC) for $0.008 \leq Z_{\rm i} \leq 0.02$. The SMC is slightly more metal-rich than WLM \citep[$12+\log({\rm O/H}) \sim$ 8--8.4, compared to 7.8 for WLM;][]{Lee+2005, Toribio+2017}, and it is as yet unclear how strongly this metallicity difference affects the AGB colors (see \S\ref{sec:metals}). The simulation is scaled to match the total number of AGB stars in the combined HST$+$JWST footprint (Fig.~\ref{fig:map}), and is shown in the lower panels of Figure~\ref{fig:jwstccd} with large symbols. The boundary between C and M stars remains similar to the boundary derived using the HST classification (Eqs.~\ref{eq:250} and \ref{eq:430}).  


The lower panels in Figure~\ref{fig:jwstccd} also show the same simulation for 10$\times$ WLM's stellar mass to highlight the more sparsely populated regions of the CCD (C- and M-type stars are marked with light red and blue colors). This more massive simulation shows that the JWST color cuts devised using the HST classifications may miss a non-negligible number of C stars in a more populated system. In fact, the massive simulation suggests that the F430M color (panels B and D) may be the better choice for classifying AGB stars overall. See \S\ref{sec:disc} for more discussion.

Given WLM's low stellar mass and sparsely populated M star branch, it is difficult to know exactly how much overlap occurs between C-type and M-type stars at metallicities lower than the SMC. If we look at the two HST-confirmed M-type stars with the deepest H$_2$O absorption (marked with large black squares in Fig.~\ref{fig:jwstccd}), it appears as though the M star branch might curve upwards in the F430M CCD, creating almost total overlap with the C star branch. We discuss this point further in \S\ref{sec:metals}. The two stars marked with circles that also follow this upwards trajectory are also HST-confirmed M-type stars, though they appear to have less extreme H$_2$O absorption, and may be red in F150W$-$F430M due to circumstellar dust (see \S~\ref{sec:dust}). There is one star that is not in the HST footprint that appears to follow the simulated M star branches well (marked with a large black triangle), which may indicate that a more populated galaxy would have more stars along this branch. On the other hand, this star is a possible contaminant (\S\ref{sec:contam}).  A larger JWST footprint in WLM and/or observations in similarly metal-poor systems would help to verify the behaviour of metal-poor M type stars.

%

\subsubsection{Isolating Massive Red Supergiants}

Figure~\ref{fig:jwstccd} also shows several stars brighter than the TRGB that are not classified as TP-AGB stars in either the HST data or the simulations (black/grey points with F090W--F150W $\lesssim$ 1.3~mag). This population includes a combination of massive RSGs (also referred to as red helium-burning -- or RHeB -- stars), main sequence (MS) stars, blue helium burning (BHeB) stars, and foreground stars. 

\begin{figure}
\centering
\includegraphics[width=0.9\columnwidth]{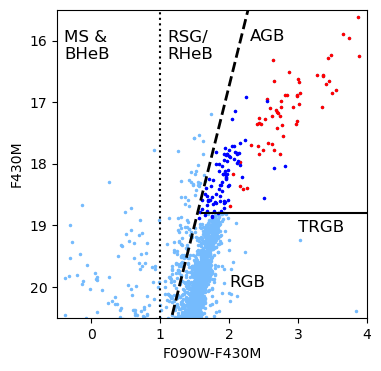}
\caption{F090W$-$F430M vs.\ F430M CMD showing the best separation of RSGs and AGB stars in the ERS filters (dashed line; Eq.~\ref{eq:rsg}). Points to the left of the dotted line are mostly likely MS stars or BHeB stars. AGB stars are color-coded according to Equation~\ref{eq:250} (C-type stars are red, M-type are dark blue).
\label{fig:rsg}}
\end{figure}

We find that RSGs are most easily isolated in the F090W-F430M vs.\ F430M color-magnitude diagram (Fig.~\ref{fig:rsg}), where they form a distinct branch blueward of the M-type AGB stars, similar to what is seen in other near-IR surveys of nearby star-forming galaxies \citep[e.g.,][]{Nikolaev+2000, Blum+2006, Boyer+2011, Rosenfield+2016}. We follow \citet{Rosenfield+2016} to define a boundary for the RSGs that depends on the color of the TRGB, which is itself dependent on the metallicity of the galaxy:

\begin{equation}
\begin{split}
    M_{\rm F430M} = -7.4 - 4.5[&(M_{\rm F090W} - M_{\rm F430M}) - \\
    &(M_{\rm F090W}^{\rm TRGB} - M_{\rm F430M}^{\rm TRGB})];
\end{split}
\label{eq:rsg}
\end{equation}

\noindent where we assume a true distance modulus of $\mu = 24.93$~mag \citep{Albers+2019}. 
Note that this boundary is conservative, in that it minimizes the loss of M-type AGB stars by keeping the dividing line closer to the RSG branch.  It is, however, likely that a small handful of RSGs fall in our M-type AGB sample. Using Equation~\ref{eq:rsg}, we find 33 RSGs that are brighter than either the F150W or F250M TRGB.

Points blueward of F090W-F430M = 1~mag are classified as MS and BHeB stars. Foreground stars can also populate this color-magnitude space, though NIRCam's small FOV and WLM's Galactic latitude ($b=-73.6^{\circ}$) together result in minimal foreground in our sample. 

\begin{figure*}
\centering
\includegraphics[width=0.7\textwidth]{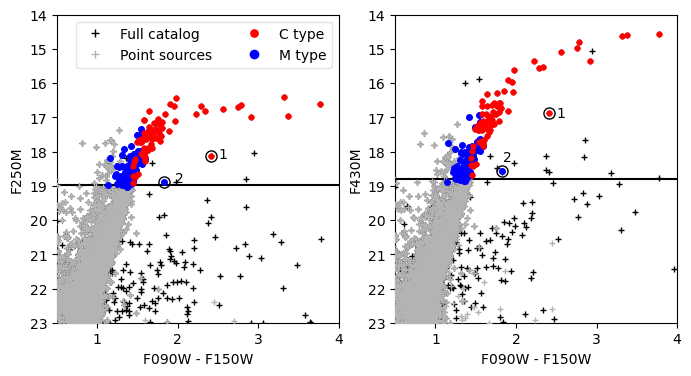}
\caption{JWST CMDs showing the extended objects that were cut from the final point-source catalog in black. Most of these are likely background objects or very red objects belonging to WLM, such as YSOs. Without the quality cuts, a handful of these would be in our AGB sample since they are brighter than the F250M TRGB (black line). Even with the quality cuts, two objects remain in our sample, circled in black and marked 1 and 2.
\label{fig:contam}}
\end{figure*}

\subsubsection{Contamination}
\label{sec:contam}

\begin{figure}
\centering
\includegraphics[width=\columnwidth]{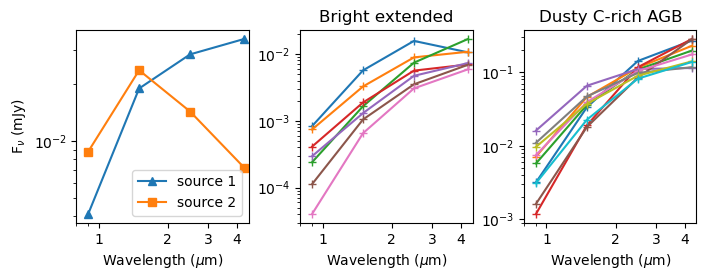}
\caption{{\em Left:} SEDs of the two sources highlighted in Figure~\ref{fig:contam}. {\em Middle:} SEDs of objects brighter than the F430M TRGB that were identified as extended based on {\tt DOLPHOT} quality cuts. {\em Right:} SEDs of the brightest and reddest C-type AGB stars in our sample. 
\label{fig:seds}}
\end{figure}

To produce our final point-source catalog, we applied several quality cuts to eliminate objects that are slightly extended based on the DOLPHOT {\em sharp, crowd, flag,} and {\em object type} parameters \citep{Warfield+2023, Weisz+2024}. Figure~\ref{fig:contam} shows that a large fraction of the eliminated extended objects are redward of the RGB\@. The Spitzer surveys of the SMC and LMC \citep{Meixner+2006, Gordon+2011} showed that these objects are dominated by unresolved background galaxies, young stellar objects (YSOs), and rare objects such as planetary nebulae \citep{Boyer+2011, Ruffle+2015, Jones+2017}. While our quality cuts eliminated the vast majority of these objects, two remain in this region of the CMD that are classified as AGB stars based on their colors/brightnesses, marked as 1 and 2 in Figure~\ref{fig:contam}. Given their positions below the extension of the C-type AGB branch in Figure~\ref{fig:contam}, their status as AGB stars is suspect. Figure~\ref{fig:seds} shows their JWST spectral energy distributions (SEDs). Source 2 peaks near 1~$\mu$m, and Source 1 peaks somewhere past 4.3~$\mu$m. Based on their SED shapes and positions in the CMD, these could be AGB stars, or perhaps post-AGB stars, given that their SEDs are similar to those in Fig.~6 from \citet{Ruffle+2015}. Spitzer observations did not detect any variability \citep{Boyer+2015, Goldman+2019}. Longer wavelength data and/or spectroscopy is needed to confirm their nature. 

By requiring our TP-AGB and RSG candidates to be brighter than the F250M TRGB, most interlopers are eliminated even without the {\tt DOLPHOT} quality cuts. However, we note that several of these slightly extended objects are brighter than the TRGB at F430M (Fig.~\ref{fig:contam}), and would be classified primarily as C-rich AGB stars if they remained in our sample. We show their SEDs in Figure~\ref{fig:seds}, compared to some example dusty C-type AGB stars (F090W$-$F150W $>$ 2) from this dataset. The SEDs of the two populations are mostly indistinguishable at these wavelengths, and we cannot eliminate the possibility that objects in this region of the CMD are evolved stars. However, since they fall below the primary C-type AGB branch at F430M = 14--16~mag, it is likely that if they belong to WLM, they are YSOs, planetary nebulae, or post-AGB stars.  We exclude these extended objects from our sample, but include sources 1 and 2 identified in Figure~\ref{fig:contam}.

\begin{deluxetable}{ccccc}
\tablecaption{Number of Evolved Stars\label{tab:estars}}
\tablehead{
\colhead{Classification} & \colhead{Field} & \colhead{$N_{\rm M}$} & \colhead{$N_{\rm C}$} & \colhead{$N_{\rm RSG}$}}
\startdata 
JWST & North NIRCam  & 61 & 40 & 24 \\
JWST & South NIRCam  & 26 & 30 & 9 \\
HST &HST FOV & \HM & \HC & \nodata \\
HST & NIRCam FOV & \JHM & \JHC & \nodata\\ 
\enddata
\tablecomments{\ For the JWST classifications, the AGB stars (C- and  M-type) are based on Equation~\ref{eq:250} and the RSGs are based on Equation~\ref{eq:rsg}. In both cases, all stars are brighter than either the F150W or F250M TRGB.}
\end{deluxetable}



\section{Discussion}
\label{sec:disc}

\subsection{Final Evolved Star Selection}
\label{sec:final}

\begin{figure}
\plotone{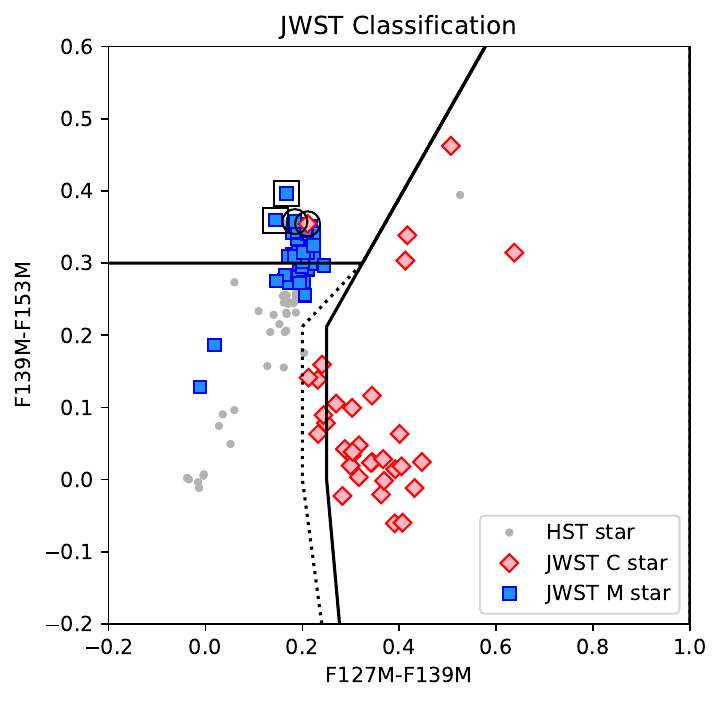}
\caption{JWST-classified TP-AGB stars (Eq.~\ref{eq:250}) shown on the HST CCD. The large squares and circles indicate the stars marked with the same symbols in Figure~\ref{fig:jwstccd}. Only one star conflicts with the HST classification: the red diamond in the M star region of the CCD.
\label{fig:hstccd}}
\end{figure}

Based on the HST classifications and the PARSEC-COLIBRI models, our preferred selections are Equation~\ref{eq:250} for TP-AGB stars and Equation~\ref{eq:rsg} for RSGs, resulting in the statistics listed in Table~\ref{tab:estars}. This selection may minimize overlap between C stars and late-type M stars. Out of a total of 157 TP-AGB stars in WLM, only two are outliers and possibly contaminants based on their JWST colors (stars 1 and 2 in \S\ref{sec:contam}). Only one star conflicts with the HST classification, a carbon star marked with a black circle in Figures~\ref{fig:jwstccd} and \ref{fig:hstccd}. This star sits right on the edge of our JWST-based classification, but is likely an M-type star based on its HST colors. The classification based on F430M (Eq.~\ref{eq:430}) mis-identifies 4 stars: 3 M-type and 1 C star.

Our JWST classification scheme has also picked up 16 more M-type stars in the HST footprints with $\rm{F139M-F153M} < 0.3$ (Fig.~\ref{fig:hstccd}). Two of these are clearly outliers and are likely contaminants, based on their positions in the HST CCD \citep[see][]{Boyer+2017}.  According to the PARSEC-COLIBRI models, the rest of the stars in this region (about 30\% of the total M star sample) are a mix of upper RGB stars and early-AGB stars. Based on their effective temperatures, they may be K-type stars, but they cannot be excluded from our M-star sample with our set of JWST filters. 

We can use the ratio of C to M stars (C/M) to compare the HST and JWST classifications, which differ mostly by the additional M/K-type stars picked up by the JWST classification. The C/M ratio is sometimes used as a proxy for metallicity \citep[e.g.,][]{Groenewegen2006b}. Here, we use the relation between C/M and the metallicity derived by \citet{Boyer+2019} and assume a metallicity of ${\rm [M/H]} = -1.27 \pm 0.04$, measured using medium resolution spectroscopy of the Calcium II Triplet equivalent widths in RGB stars \citep{Leaman+2009}. Within the combined HST+JWST footprints (Fig.~\ref{fig:map}), the HST and JWST-derived C to M ratios are ${\rm (C/M)_{HST}} = 0.91\pm 0.16$ and ${\rm (C/M)_{JWST}} = 0.68\pm0.12$, yielding [M/H] values of $-1.23\pm0.07$ for HST and $-1.11\pm0.07$ for JWST. The HST value is clearly closer to the metallicity measured by \citet{Leaman+2009}. However, low number statistics are playing a role here, since the C/M value in the entire HST footprint, derived using the HST classifications is $0.80\pm0.10$, while the overall JWST C/M in the entire NIRCam footprint is {\em also} $0.80\pm0.10$. These values yield a metallicity of ${\rm [M/H]} = -1.18\pm0.05$, within one sigma of the Leaman value. 

Moreover, all of these C/M values are within the uncertainties predicted from C/M values from several nearby dwarf galaxies, which were derived using the HST classification scheme and an optical classification technique that uses narrow band CN and TiO filters \citep[see Fig.~9 from][]{Boyer+2019}. We conclude that while the HST and JWST classification schemes give slightly different results, the differences result primarily from low number statistics.

\begin{figure}
\centering
\includegraphics[width=1\columnwidth]{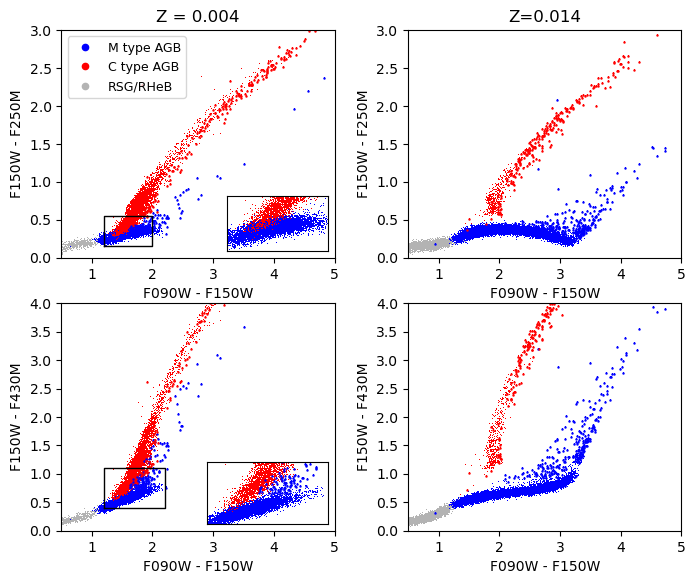}
\caption{Simulations using a constant star-formation rate at two metallicities, showing only stars brighter than the TRGB. The insets in the left hand panels are zoomed into the region outlined by a black box. At low metallicity (left), it is difficult to isolate the bluest C stars with F250M, and difficult to isolate dusty M-type AGB stars with F430M. At high metallicity (right), either filter works across the entire F090W$-$F150W color range. 
\label{fig:constsfr}}
\end{figure}

\subsection{Metallicity Effect on TP-AGB colors}
\label{sec:metals}

As noted in \S\ref{sec:jwstclass}, the COLIBRI models that we compare to here are calibrated to the Magellanic Clouds, which are more metal-rich than WLM. In this section, we investigate the effect metallicity has on the models and on the JWST colors. Figure~\ref{fig:constsfr} shows simulations from COLIBRI+TRILEGAL, this time with a constant SFH to ensure all TP-AGB ages are well populated. We simulated two metallicities, one at SMC metallcity ($Z=0.004)$, and one at high metallicity ($Z=0.014$, about 70\% solar). The higher-metallicity COLIBRI models are currently calibrated to the LMC \citep{Pastorelli+2020}, at $Z=0.008$. The most dramatic differences between these simulations is the color of the M-type stars on the x-axes and the bluest extent of the C stars. These differences are presumably caused by the cooler temperatures and deeper molecular absorption at high metallicity, as well as the narrower age range of C stars at high metallicity. 

As the metallicity increases, the separation between C- and M-type stars improves significantly. At SMC metallicity, the different stellar types have quite a narrow separation, made even narrower by the presence of many {\em dusty} M-type stars, which appear to be rare in WLM. This metallicity dependence suggests that in lower metallicity systems like WLM, we should expect even more overlap between the C- and M-type stars. Indeed, this does appear to be the case, as illustrated in the upper panels of Figure~\ref{fig:jwstccd}. However, it is difficult to discern how severe the overlap is, given the small number of late-type M stars in WLM. In a more populated metal-poor system, it is possible that the JWST colors used by this ERS program would not provide a good separation between AGB subtypes. On the other hand, late-type M stars are quite rare in metal-poor systems in general, so this overlap may not have a significant effect on the overall statistics. Nevertheless, we advise caution when using the ERS filters (F090W, F150W, F250M, and F430M) to classify AGB stars in very metal-poor systems.

\subsubsection{Other JWST Filters for Evolved Star Classification}
\label{sec:others}

\begin{figure}
\centering
\includegraphics[width=1\columnwidth]{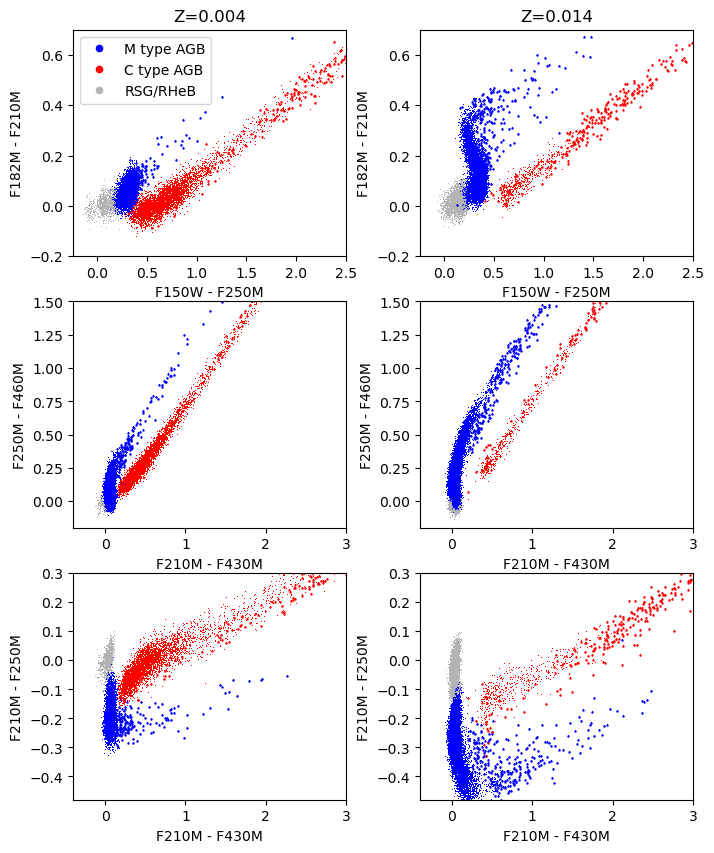}
\caption{Additional JWST filters that show good separation between AGB subtypes, according to the same constant star-formation rate simulation shown in Figure~\ref{fig:constsfr}. For reference, these filters are also plotted in Figure~\ref{fig:filters}. The F210M$-$F430M shows the widest separation between the bulk of the C- and M-type AGB stars. The combinations shown in the upper and lower panels also show good isolation of the RSGs (grey). Only stars brighter than the TRGB are plotted here.
\label{fig:others}}
\end{figure}

Other JWST filters sample interesting features in TP-AGB and RSG stars. For example, F140M, F182M, and F210M sample H$_2$O, F300M, F335M, F410M, and F430M sample HCN$+$C$_2$H$_2$. Despite this, we were unable to find filter combinations that effectively isolate AGB stellar types from one another, independent of metallicity. Using the simulations with constant star formation rate described above, we tested every possible combination, and found that most combinations are either worse or similar to the ERS filters in their classification effectiveness. Only a handful of options show some marginal improvement, shown in Figure~\ref{fig:others}. All of these options include F210M and/or F430M (see Fig.~\ref{fig:filters}). The F210M$-$F430M color specifically has the widest separation between the bulk of the M-type stars and C-type stars, though circumstellar dust can complicate matters. The upper and lower panels in Figure~\ref{fig:others} show combinations that are also quite good at separating RSGs (in grey) from the TP-AGB stars.  

As with the ERS filters discussed in \S\ref{sec:metals}, it is possible that the C and M stars will show more overlap at lower metallicities. However, the HST medium-band filters appear to work well over a wide range of metallicity \citep{Boyer+2017, Boyer+2019}. If a clean sample of these stellar types in a metal-poor galaxy is essential, the HST medium-band WFC3/IR filters may be the better option, despite HST's decreased sensitivity and angular resolution in the near-IR. We are providing the simulations shown in Figure~\ref{fig:others} along with this manuscript, in every NIRCam filter, to assist in future JWST program designs.\footnote{{\color{red} We plan to post the simulations to CDS and will link to them in this footnote.}}

\subsection{Dust}

\label{sec:dust}




With JWST, we find 10 very dusty AGB star candidates in WLM (F150W$-$F430M $>$ 2~mag), and many more slightly dusty candidates (F150W$-$F430M = 0.7--2~mag; Fig.~\ref{fig:jwstcmds}).  With Spitzer, contamination from unresolved background galaxies made it impossible to identify dusty objects belonging to WLM, so Spitzer surveys relied on variability information with 2--8 epochs of data \citep{Boyer+2015, Goldman+2019}. With just two epochs of data, \citet{Boyer+2015} were able to identify 9/10 of the JWST very dusty AGB stars.  However, even with 8 epochs, \citet{Goldman+2019} could not identify any of the less-dusty AGB stars, presumably because their pulsation amplitudes are too small. JWST data are thus clearly essential to obtaining a full picture of the dust budget in galaxies more distant than the Magellanic Clouds.

All of the dusty stars with F150W$-$F430M $>$ 2~mag are identified as carbon stars using the classification described in \S\ref{sec:final}. Of the less dusty stars, 11\% are M-type. While we cannot measure the dust mass produced by these stars without data beyond 5~\micron, it is clear that the present-day dust input by evolved stars in WLM is dominated by carbonaceous dust, and thus by stars with masses between about 1.5--3~$M_\odot$ \citep{Marigo+2013}.  

Dusty massive {\em M-type} stars such as those seen in the extremely metal-poor galaxies Sextans\,A and Sag\,DIG \citep{Boyer+2017} are of particular interest to the evolution of dust in the early Universe since they can start producing dust as early as 30~Myr after they form. However, based on their brightnesses, the most massive dusty M-type AGB stars that we see in WLM are likely $\sim$2.5--3.5~$M_\odot$, which corresponds to a dust input timescale of about 500 Myr -- too late to be proxies of dusty AGB stars in the early Universe. Based on their colors and magnitudes, it is also possible that two of the dusty M-type AGB candidates are actually RSGs, which is possible because WLM has ongoing star formation and evidence for very recent star formation \citep{mondal18}. If so, that would put their masses between about 8 and 23~$M_\odot$, and currently existing dust would be at least partially destroyed by their future supernova explosions.


\section{Conclusions}

We imaged WLM in four NIRCam filters as part of the JWST Resolved Stellar Populations ERS program. These filters were selected, in part, to investigate how effectively different subtypes of evolved stars can be identified with JWST, specifically TP-AGB stars and RSGs. In a single NIRCam footprint, we find 70 carbon TP-AGB stars, 87 M-type TP-AGB stars, and 33 RSGs. Of these, 72 (or about 46\%) show evidence of circumstellar dust.

We find that the ERS filters (F090W, F150W, F250M, and F430M) are excellent for isolating RSGs, but the efficacy of the AGB selection is likely highly dependent on metallicity. A comparison to the well-established HST medium-band AGB classification strategy shows very good agreement to the ERS classification defined here for WLM, resulting in an identical C/M ratio derived from both techniques. This C/M ratio (0.8$\pm$0.1) follows the relationship of C/M versus metallicity derived using HST medium-band data of other nearby dwarf galaxies.  

However, WLM is a low-mass system and short-lived stellar types such as late-type M AGB stars and dusty AGB stars are rare. Metal-poor systems with more massive populations may suffer from significant overlap between these rarer objects and carbon-rich AGB stars using these filters. The COLIBRI stellar evolution models suggest that the ERS filters will be more effective in more metal-rich systems, where cooler temperatures and a narrowing mass range for carbon stars leads to a much wider separation of AGB subtypes.  We also investigated other NIRCam filters and found that combinations using F182M, F210M, and F460M may provide some improvements at low metallicity.

We also compared WLM's dusty AGB population to that identified with Spitzer by \citet{Boyer+2015}. We find that Spitzer was only able to identify the 9 most extreme examples out of 71 total dusty AGB stars in the NIRCam footprint. Of these dusty stars, 89\% are carbon-rich, confirming that carbonaceous dust dominates the dust input into WLM's interstellar medium. While dusty objects are typically studied in the mid-IR, these results demonstrate NIRCam's ability to efficiently identify dust-producing stars photometrically, while also providing information on the basic dust species (carbonaceous or siliceous). NIRCam is thus a valuable tool for the identification of dust-producing stars in much more distant systems than can be reached by JWST's Mid-InfraRed Instrument (MIRI), which has lower spatial resolution and sensitivity.

\begin{acknowledgments}
This work is based in part on observations made with the NASA/ESA/CSA JWST. The data were obtained from the Mikulski Archive for Space Telescopes at the Space Telescope Science Institute, which is operated by the Association of Universities for Research in Astronomy, Inc., under NASA contract NAS 5-03127 for JWST. These observations are associated with program ERS-1334.  This research is also based in part on observations made with the NASA/ESA Hubble Space Telescope obtained from the Space Telescope Science Institute, which is operated by the Association of Universities for Research in Astronomy, Inc., under NASA contract NAS 5–26555. These observations are associated with program HST-GO-16162. Support for this work was provided by NASA through grants ERS-1334 from the Space Telescope Science Institute, which is operated by AURA, Inc., under NASA contract NAS5-26555. This research has made use of NASA Astrophysics Data System Bibliographic Services and the NASA/IPAC Extragalactic Database (NED), which is operated by the Jet Propulsion Laboratory, California Institute of Technology, under contract with the National Aeronautics and Space Administration. E.N.K.\ acknowledges support from NSF CAREER grant AST-2233781. We thank the JWST teams at STScI for a smooth execution of this ERS program and for assistance in its design and calibration. 
\end{acknowledgments}

\vspace{5mm}
\facilities{HST(WFC3), JWST(NIRCam)}

\software{Astropy \citep{astropy:2013, astropy:2018, astropy:2022},  
          \texttt{DOLPHOT} \citep{Dolphin2000, Dolphin2016}, Scipy \citep{scipy2020},  TRILEGAL \citep{Girardi+2005}
          }



\bibliography{main}{}
\bibliographystyle{aasjournal}

\end{document}